\documentclass[preprint,12pt]{elsarticle}

\usepackage{amssymb}
\usepackage{amsmath}
\usepackage{multirow}
\usepackage{url}
\usepackage[table,xcdraw]{xcolor}
\usepackage[hidelinks]{hyperref}
\usepackage{footmisc}
\usepackage{graphicx}

\begin{document}

\begin{frontmatter}



\title{ViTaL: A Multimodality Dataset and Benchmark for Multi-pathological Ovarian Tumor Recognition}


\author[Beihang]{You Zhou\fnref{equal}} 
\ead{sy2402322@buaa.edu.cn}
\author[Beihang]{Lijiang Chen\fnref{equal}} 
\ead{chenlijiang@buaa.edu.cn}
\author[shijitan]{Guangxia Cui\fnref{equal}}
\ead{cuigx@bjsjth.cn}
\author[shijitan]{Wenpei Bai\corref{corresponding}}
\ead{baiwp@bjsjth.cn}
\author[shijitan]{Yu Guo}
\ead{1923197447@mail.ccmu.edu.cn}
\author[Beihang]{Shuchang Lyu\corref{corresponding}}
\ead{lyushuchang@buaa.edu.cn}
\author[liverpool]{Guangliang Cheng}
\ead{Guangliang.Cheng@liverpool.ac.uk}
\author[Beihang]{Qi Zhao}
\ead{zhaoqi@buaa.edu.cn}

\affiliation[Beihang]{organization={Department of Electronics and Information Engineering, Beihang University},
            addressline={Xueyuan Road No.37}, 
            city={Haidian district},
            postcode={100191}, 
            state={Beijing},
            country={China}}
\affiliation[shijitan]{organization={Department of Gynecology and Obstetrics, Beijing Shijitan Hospital, Capital Medical University},
            addressline={Tieyi Road No.10}, 
            city={Haidian district},
            postcode={100038}, 
            state={Beijing},
            country={China}}
\affiliation[liverpool]{organization={Department of Computer Science, University of Liverpool},
            addressline={Foundation Building, Brownlow Hill}, 
            city={Liverpool},
            postcode={L693BX},
            country={UK}}
            
\cortext[corresponding]{Corresponding authors}
\fntext[equal]{Contribute Equally.}
\begin{abstract}
Ovarian tumor, as a common gynecological disease, can rapidly deteriorate into serious health crises when undetected early, thus posing significant threats to the health of women. Deep neural networks have the potential to identify ovarian tumors, thereby reducing mortality rates, but limited public datasets hinder its progress. To address this gap, we introduce a vital ovarian tumor pathological recognition dataset called \textbf{ViTaL} that contains \textbf{V}isual, \textbf{T}abular and \textbf{L}inguistic modality data of 496 patients across six pathological categories. The ViTaL dataset comprises three subsets corresponding to different patient data modalities: visual data from 2216 two-dimensional ultrasound images, tabular data from medical examinations of 496 patients, and linguistic data from ultrasound reports of 496 patients. It is insufficient to merely distinguish between benign and malignant ovarian tumors in clinical practice. To enable multi-pathology classification of ovarian tumor, we propose a ViTaL-Net based on the Triplet Hierarchical Offset Attention Mechanism (THOAM) to minimize the loss incurred during feature fusion of multi-modal data. This mechanism could effectively enhance the relevance and complementarity between information from different modalities. ViTaL-Net serves as a benchmark for the task of multi-pathology, multi-modality classification of ovarian tumors. In our comprehensive experiments, the proposed method exhibited satisfactory performance, achieving accuracies exceeding 90\% on the two most common pathological types of ovarian tumor and an overall performance of 85\%. Our dataset and code are available at https://github.com/GGbond-study/vitalnet.
\end{abstract}




\end{frontmatter}


\section{Introduction}
\label{sec:introduction}
Ovarian tumor, as a significant threat to women's health, have witnessed a marked increase in both incidence and mortality rates in recent years, emerging as a crucial issue in global public health~\cite{rong2024decision}. According to Global Cancer Statistics 2022~\cite{bray2024global}, in many transitioning countries, the mortality rate of ovarian tumor among women remains high, second only to that of breast cancer and cervical cancer. Due to the lack of early symptoms and reliable screening tests, most ovarian tumors are diagnosed at advanced stages, severely impacting patients' health and quality of life~\cite{fahim2024ovanet}. Early detection is crucial for reducing mortality~\cite{yang2023triple}. Current diagnostic methods include pelvic exams, imaging tests, blood tests, tumor marker tests like CA-125, and genetic tests~\cite{zhao2022mmotu}. These processes require skilled clinicians, making computer-aided diagnosis of ovarian tumors both urgent and meaningful.
\par
Artificial intelligence (AI) has seen significant growth after 2010 with the emergence of deep learning (DL) using neural networks. DL methods have been extensively studied for their applications in medical image analysis~\cite{tang2025medical}, spanning a wide range of clinical areas including neurology~\cite{meidani2022development}, retinal imaging~\cite{xu2025joint}, pulmonology~\cite{idrisoglu2024copdvd}, digital pathology~\cite{bidgoli2022evolutionary}, breast imaging~\cite{chereda2024stable}, cardiology~\cite{jeong2023artificial}, abdominal imaging~\cite{hao2022self}, and musculoskeletal imaging~\cite{tang2024follow}. These applications have demonstrated the potential of DL to enhance diagnostic accuracy, improve patient outcomes, and streamline clinical workflows across various medical specialties. In recent years, some researches~\cite{zhao2022mmotu} focus on analyzing two-dimensional images of tumors, such as 2D ultrasound images, using computer-aided methods to attempt to construct algorithmic models for recognizing between benign and malignant ovarian tumors. During clinical examinations, in addition to common two-dimensional imaging tests, physicians typically also arrange for the detection of tumor marker CA125~\cite{kim2023prognostic}. This means that in clinical practice, it may be relatively possible to obtain multi-modality information from patients, including two-dimensional ultrasound image data, image report and tumor marker information. These multi-modality information reflect the biological characteristics and pathological features of ovarian tumors from different perspectives, providing a rich data basis for more accurate diagnosis of ovarian tumors. Existing research methods~\cite{samaee2025multi,tan2022multi,zhang2024scalable} do not effectively integrate these multi-modality information, resulting in a large amount of valuable data in clinical practice not being fully utilized, and thus making it difficult to achieve precise recognition of the benignity and malignancy of ovarian tumors. Despite their notable works, there still exists the following two main weaknesses. First of all, due to the limitations of public ovarian tumor datasets, many studies~\cite{wanderley2019end,wu2018deep,li2019cr} primarily focus on single-modality recognition and diagnosis, and their diagnosis only distinguishing between benign and malignant tumors. Secondly, recent notable studies~\cite{li2024review} attempted to integrate information from multiple modalities by a decision-level fusion module to achieve precise recognition of tumors. This simple and rigid fusion can lead to significant loss when integrating information from different modalities. 
\par
To address these two weaknesses, we first construct an ovarian tumor dataset called ViTaL containing visual, tabular and linguistic data of 496 cases. ViTaL dataset consists of three subsets, each corresponding to different modalities of patient information: visual information of 2216 two-dimensional ultrasound images, tabular information of medical examinations of 496 patients, and linguistic information of 496 patients' ultrasound reports. Based on ViTaL dataset, we conduct a mutli-modality, multi-pathological classification task for ovarian tumor. This task involves using information from visual, tabular and linguistic modalities data to perform precise classification of six specific types of ovarian tumors, rather than the coarse classification of benign and malignant. To achieve better classification performance, we propose the ViTaL-Net which is capable of processing visual, tabular, and linguistic data simultaneously. It is mainly accomplished through the Triplet Hierarchical Offset Attention Mechanism (THOAM)  to extract and integrate information from different modalities of data, which lies in enhancing the correlation and complementarity between information extracted from different modalities of data through attention mechanism. By introducing triplet hierarchical Offset attention mechanism (THOAM), the proposed method can fully mine and explore useful multi-modality information, improving classification performance for ovarian tumor.
\par
This method is designed to fully harness the potential of clinical multi-modality information and build a more precise and effective model for recognizing ovarian tumors. In this way, it provides a reliable basis for clinical diagnosis and treatment decision-making. To sum up, the main contributions are as follows:
\begin{itemize}
    \item We constructed an ovarian tumor dataset named ViTaL, which includes 496 cases of patient visual, tabular and linguistic information. Each case contains their corresponding two-dimensional ultrasound images, reports and tabular examination information.
    \item To the best of our knowledge, we are the first to conduct precise multi-pathology classification of six specific types of ovarian tumors by using visual, tabular and linguistic data.
    \item We proposed an attention-based multi-modality fusion network called ViTaL-net. It effectively mines the information contained in each modality and reduces the loss during the fusion of features from different modalities.
    \item Extensive experiments on the proposed ViTaL dataset show that our ViTaL-Net outperforms previous state-of-the-art networks and serves as a benchmark for ovarian tumor classification.
\end{itemize}
\par
The remainder of this paper is organized as follows. In Section 2, we review previous work relevant to our study. We introduce our proposed dataset in Section 3. Section 4 details our research methodology. Section 5 presents the experiments and results analysis. The discussion and conclusion of this work are presented in Sections 6 and 7, respectively.
\section{Related Work}
\subsection{Multi-modality medical diagnostics with computer assistance}
Tumor diagnosis relies on advanced imaging techniques such as ultrasound, mammography, CT, and MRI, which provide non-invasive insights into cellular changes and tumor environments. Recent advancements in AI and ML have significantly enhanced tumor detection by leveraging complex algorithms for precise diagnostics. U-Net~\cite{ronneberger2015u} based on Convolutional Neural Networks (CNNs) has demonstrated excellent performance in the field of medical image segmentation. ResNet~\cite{he2016deep}, with its unique residual network, has shown excellent stability and robustness in the field of medical image analysis. Medmamba~\cite{yue2024medmamba}, which combines State-Space Models (SSMs) with Convolutional Neural Networks (CNNs), has also demonstrated strong competitiveness in the field of medical classification. 
\par
There have also been numerous notable studies in the field of multi-modality medical diagnostic. Based on contrastive learning between text-image pairs, the ConVIRT~\cite{zhang2022contrastive} framework has demonstrated superior performance compared to previous methods on two public lung datasets. The GLoRIA~\cite{huang2021gloria} primarily employs attention mechanisms to match words in radiology reports with sub-regions of the image, thereby learning global-local representations of the image. The LViT~\cite{li2023lvit} leverages both CNN and Transformer architectures to process cross-modal information which demonstrates excellent performance in medical image segmentation. Most of these studies have primarily focused on the lungs and chest, with limited research on the ovaries. 
Our research is primarily dedicated to the classification of ovarian tumors by integrating data from three distinct modalities: images, text, and tables. This multimodal approach enables a more comprehensive and in-depth classification within the medical domain. Specifically, medical images provide visual information about the tumors, text data offer detailed descriptions and clinical notes, while tabular data supply structured information such as patient demographics and laboratory results. By combining these diverse types of data, we aim to capture a more complete picture of each case, thereby enhancing the accuracy and robustness of the classification outcomes. This method not only leverages the strengths of each modality but also compensates for potential limitations of any single data type, ultimately leading to a more reliable diagnostic process.
\begin{table}[t]
\caption{Statistics of existing ovarian tumor datasets and our ViTaL dataset.}
\begin{center}
\scalebox{0.68}{
    \begin{tabular}{c|cc|ccc|cc|c}
    \hline
    \multirow{2}{*}{Dataset} & \multirow{2}{*}{Year} & \multirow{2}{*}{Sample} & \multicolumn{3}{c|}{Modality}                                            & \multicolumn{2}{c|}{Task}   & \multirow{2}{*}{Category}                           \\
                             &                       &                         &Image         & Table        & Text         & Cls          & Seg          &                                                     \\ \hline
    Wu et al.~\cite{wu2018deep}                       & 2018                  & 988                     & $\checkmark$ & $\times$     & $\times$     & $\checkmark$ & $\times$     & benign, malignant and borderline                    \\
    Wanderley et al.\cite{wanderley2019end}                & 2018                  & 87                      & $\checkmark$ & $\times$     & $\times$     & $\times$     & $\checkmark$ & ovary and follicle                                  \\
    Narra et al.\cite{narra2018automated}                    & 2018                  & 1500                    & $\checkmark$     & $\times$     & $\times$     & $\times$     & $\checkmark$ & ovary and follicle                                  \\
    Li et al.\cite{li2019cr}                       & 2019                  & 3204                    & $\checkmark$ & $\times$     & $\times$     & $\times$     & $\checkmark$ & ovary and follicle                                  \\
    Wang et al.\cite{wang2021application}                 & 2021                  & 412                     & $\checkmark$  &$\times$    & $\times$     & $\checkmark$ & $\times$     & benign, malignant and borderline                    \\
    Yang et al.\cite{yang2021contrastive}                     & 2021                  & 1924                    & $\checkmark$   & $\times$     & $\times$     & $\times$     & $\checkmark$ & ovary and follicle                                  \\
    PCOS~\cite{reka2025}                      & 2021                  & 594                     & $\checkmark$   & $\times$     & $\times$     & $\checkmark$ & $\times$     & healthy and unhealthy                               \\
    MMOTU~\cite{zhao2022mmotu}                    & 2022                  & 1609                    & $\checkmark$ & $\times$     & $\times$     & $\checkmark$ & $\checkmark$ & 8 specific types of ovarian tumors \\ 
    OVATUS-V1\footnote{OVATUS-V1:\url{https://sigm-seee.github.io/datasets/Ovarian.html}}                     & 2025                  & 439                    & $\checkmark$ & $\times$     & $\times$     & $\times$ & $\checkmark$ & 6 specific types of ovarian tumors \\ \hline
    ViTaL(Ours)                    & 2025                  & 2216                    & $\checkmark$   & $\checkmark$ & $\checkmark$ & $\checkmark$ & $\times$     & 6 specific types of ovarian tumors   \\ \hline
    \end{tabular}
}
\end{center}
\label{1}
\end{table}
\subsection{Datasets For Ovarian Tumor Diagnosis}

Datasets play a pivotal role in the development of deep learning models, as these models rely heavily on large high-quality and well-annotated data to learn effectively and produce reliable results. Selecting and curating appropriate dataset is a foundational step in many machine learning projects. However, in the specific domain of ovarian tumor research, the availability of comprehensive and publicly accessible datasets is strikingly limited, as summarized in Table~\ref{1}. Many existing datasets have a limited categories of ovarian tumor which could not comprehensively cover the various types of ovarian disease. This limitation lead many notable studies on ovarian tumor recognition to primarily focus on the coarse distinction between benign and malignant tumors. Wu et al.~\cite{wu2018deep} and Wang et al.\cite{wang2021application} constructed 2D ultrasound datasets to train models that can categorize ovarian tumors into benign, borderline, and malignant. In comparison, our dataset includes six specific ovarian tumor categories. Additionally, Wanderley et al.\cite{wanderley2019end}, Li et al.\cite{li2019cr}, Narra et al.\cite{narra2018automated} and Yang et al.\cite{yang2021contrastive} created 2D or 3D ultrasound datasets for ovarian structure segmentation. Compared with the MMOTU~\cite{zhao2022mmotu} dataset, the ViTaL dataset incorporates additional data from medical examination tables and report texts. In clinical practice, it is insufficient to merely conduct a coarse classification of ovarian tumors, more accurate and detailed categorization is often required. Despite prior advanced studies attempt conduct accurate classification for ovarian pathological categories, it could not identify ovarian tumor categories more accurately from a multidimensional perspective due to limitations of datasets~\cite{hsu2022automatic, fan2023accurate}.This scarcity of data poses significant challenges for researchers aiming to develop robust and accurate diagnostic models for ovarian tumors. Therefore, building a public available high-quality dataset is crucial for advancing researches in ovarian tumor recognition.
\footnotetext{OVATUS-V1: \url{https://sigm-seee.github.io/datasets/Ovarian.html}}
\subsection{Attention Mechanisms in Medical Applications}
Attention mechanism, inspired by human visual attention, dynamically allocates resources by calculating relevance weights, focusing on the critical parts of the input data. This allows the model to prioritize important information. In 2014, Bahdanau et al. first introduced the attention mechanism~\cite{bahdanau2014neural} into neural network-based machine translation models. In 2017, Vaswani et al~\cite{vaswani2017attention} proposed the Transformer model, which introduced the self-attention mechanism and significantly improved training efficiency. The application of attention mechanisms in the medical field has made significant progress, especially in medical image recognition and disease diagnosis. Attention U-Net~\cite{oktay2018attention}, by incorporating Attention Gates, can dynamically focus on critical regions within images, thereby significantly enhancing the accuracy and precision of segmentation. Additionally, vision models based on the Transformer architecture, such as Vision Transformer (ViT)~\cite{dosovitskiy2020image}, have been widely applied in the field of medical imaging. There have also been numerous notable studies in the field of multi-modality medical diagnostic. Some studies have employed tabular data and medical imaging data for the diagnosis of Alzheimer's disease~\cite{hu2025multi}, while other studies have integrated pathological texts and medical images for retinal diagnosis~\cite{li2024text}. Most of these studies have primarily focused on the lungs and chest, with limited research on the ovaries.
\begin{figure}[t]
\includegraphics[width=\linewidth]{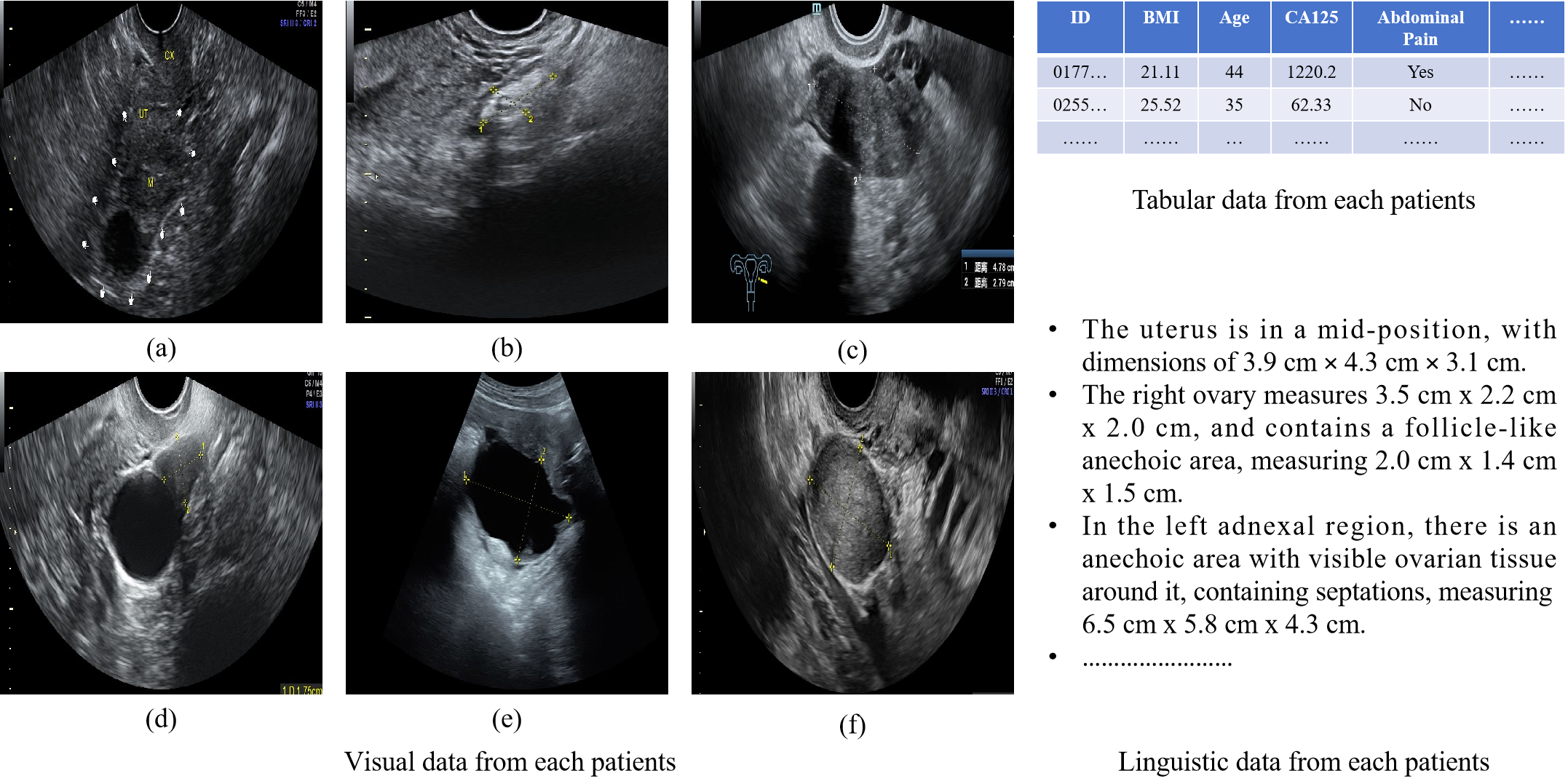}
\caption{Typical samples in ViTaL dataset, each containing visual, tabular and linguistic modality data.}
\label{example}
\end{figure}

\section{ViTaL Dataset}
The scarcity of high-quality ovarian datasets stands as a significant obstacle in the development of Computer-Aided Diagnosis systems for ovarian tumor recognition and classification. To address this critical gap, we aim to introduce a vital dataset that features comprehensive and high-quality annotations. As shown in Fig.\ref{example}, this dataset mainly includes three types of modal data: images, text, and tabular data. This dataset is intended to serve as a robust foundation for researchers and practitioners, facilitating the development of more effective Computer-Aided Diagnosis systems and fostering innovation in ovarian tumor diagnosis and treatment.
\subsection{Collection}
\begin{figure}[t]
\centering
\includegraphics[width=0.8\linewidth]{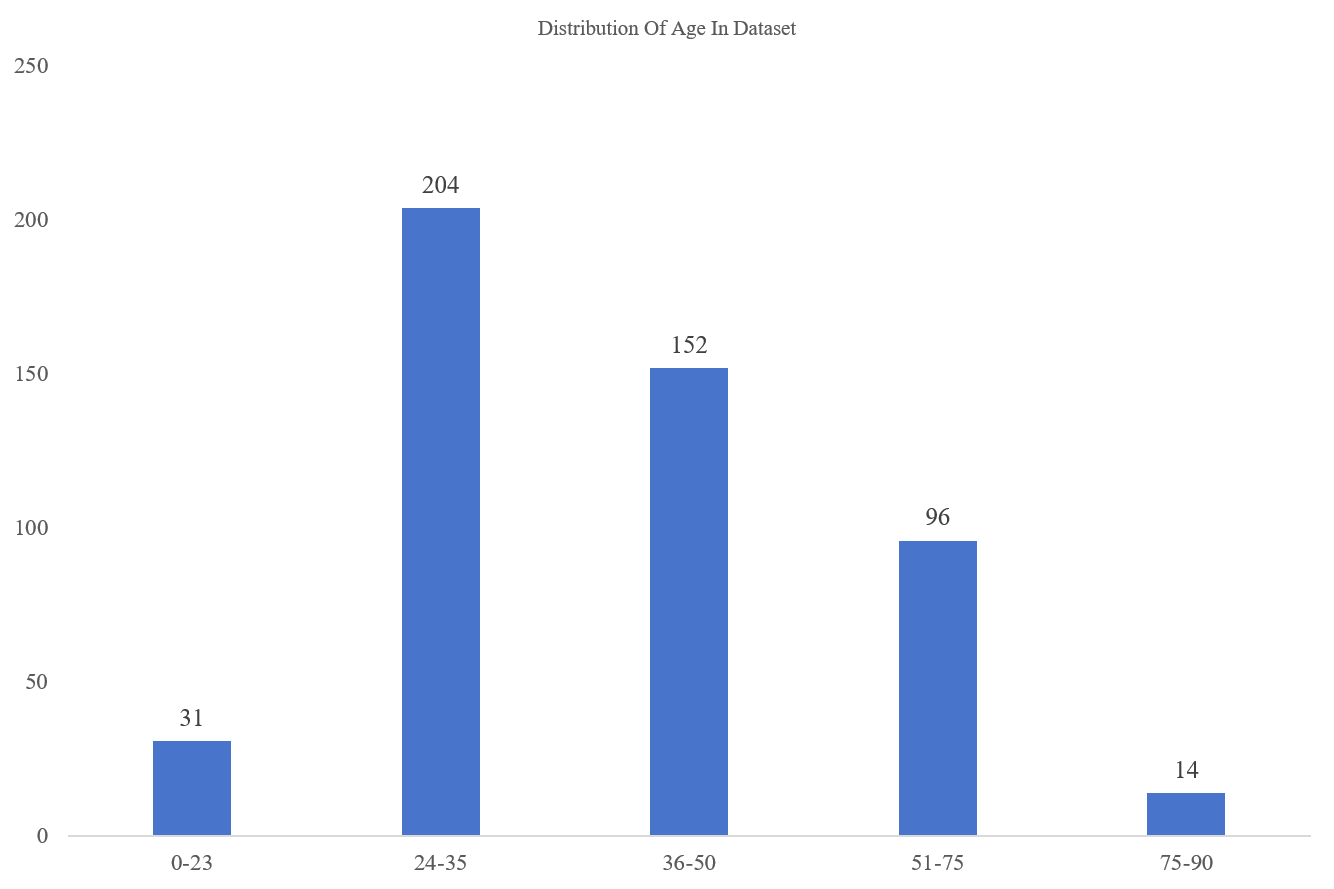}
\caption{Distribution of age in the ViTaL dataset. The horizontal axis of the table corresponds to age and the left vertical axis shows the count.}
\label{age}
\end{figure}
Several 2D and 3D open-source ovarian datasets are listed in Table~\ref{1}. Different from them, our ViTaL dataset focuses on multi-pathological classification task of ovarian tumors from data of different modalities. ViTaL dataset consists of three subsets, each corresponding to different modalities of patient information: visual information of 2216 two-dimensional ultrasound images, tabular information of medical examinations of 496 patients, and linguistic information of 496 patients' ultrasound reports. All data
are obtained from Beijing Shijitan Hospital, Capital Medical University. The scanner utilized is the Mindray Resona 8 ultrasonic diagnostic instrument. During the scanning process, doctors select specific images (slices) that clearly display the lesion regions. Since a single slice may not always provide sufficient diagnostic information, multiple scans with different viewing angles are often selected. The average age of these patients is 39.8 years.  The detailed distribution of the dataset is illustrated Fig.~\ref{age}. It is worth to note that we have collected 30 cases from patients aged 6 to 23 in the ViTaL dataset for two main reasons: First, we aim to enhance the diversity of ViTaL, improve its robustness and facilitate subsequent research. Then, we intend to improve the generalization performance of the model. When we release ViTaL, we will clearly indicate which data belong to individuals aged 18 and above. Researchers can then decide whether to incorporate this subset into their studies.
\begin{figure}[t]
\includegraphics[width=\linewidth]{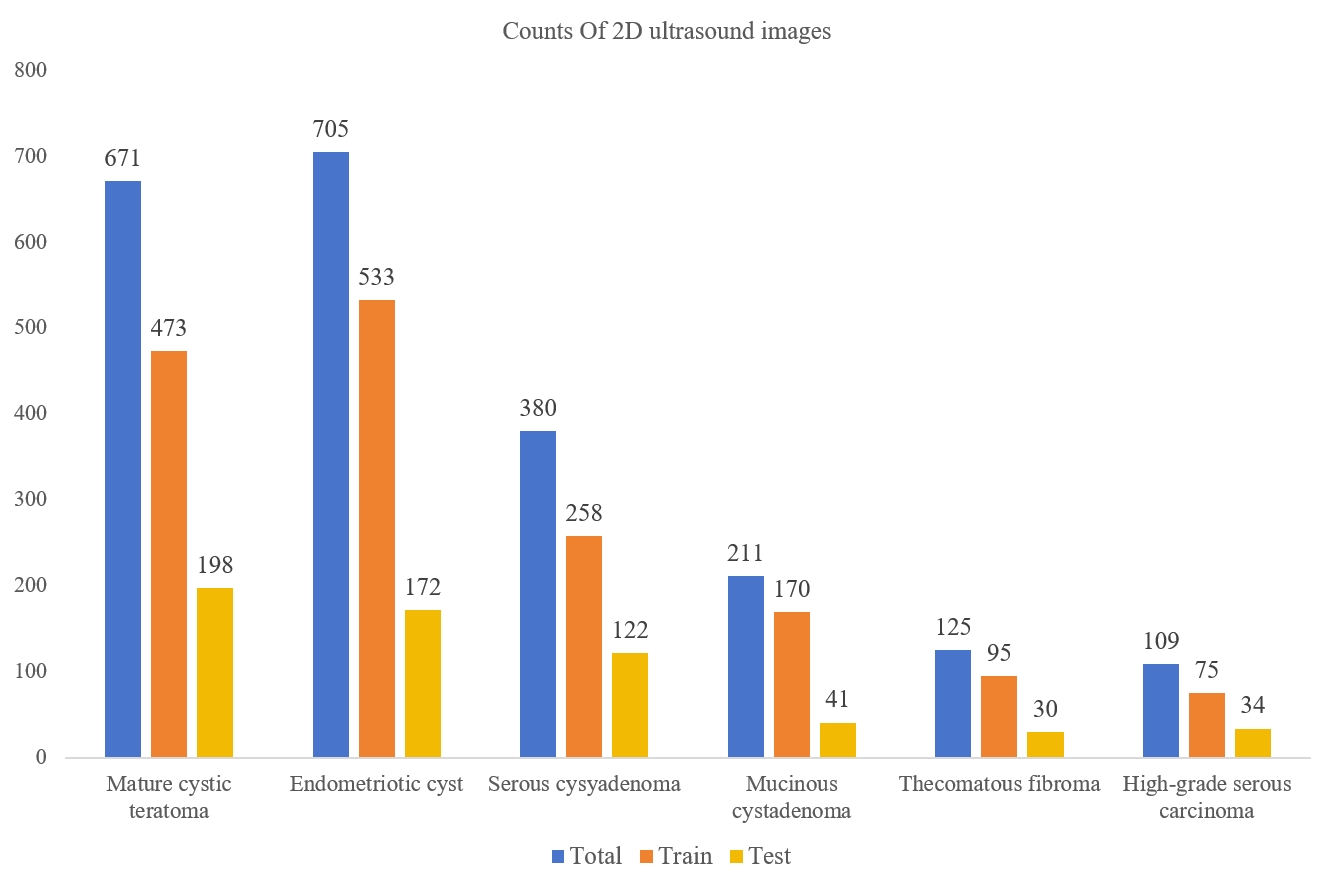}
\caption{The number of ultrasound images containing in each category. The horizontal axis of the table corresponds to category and the left vertical axis shows the count.}
\label{2d}
\end{figure}
\begin{figure}[t]
\includegraphics[width=\linewidth]{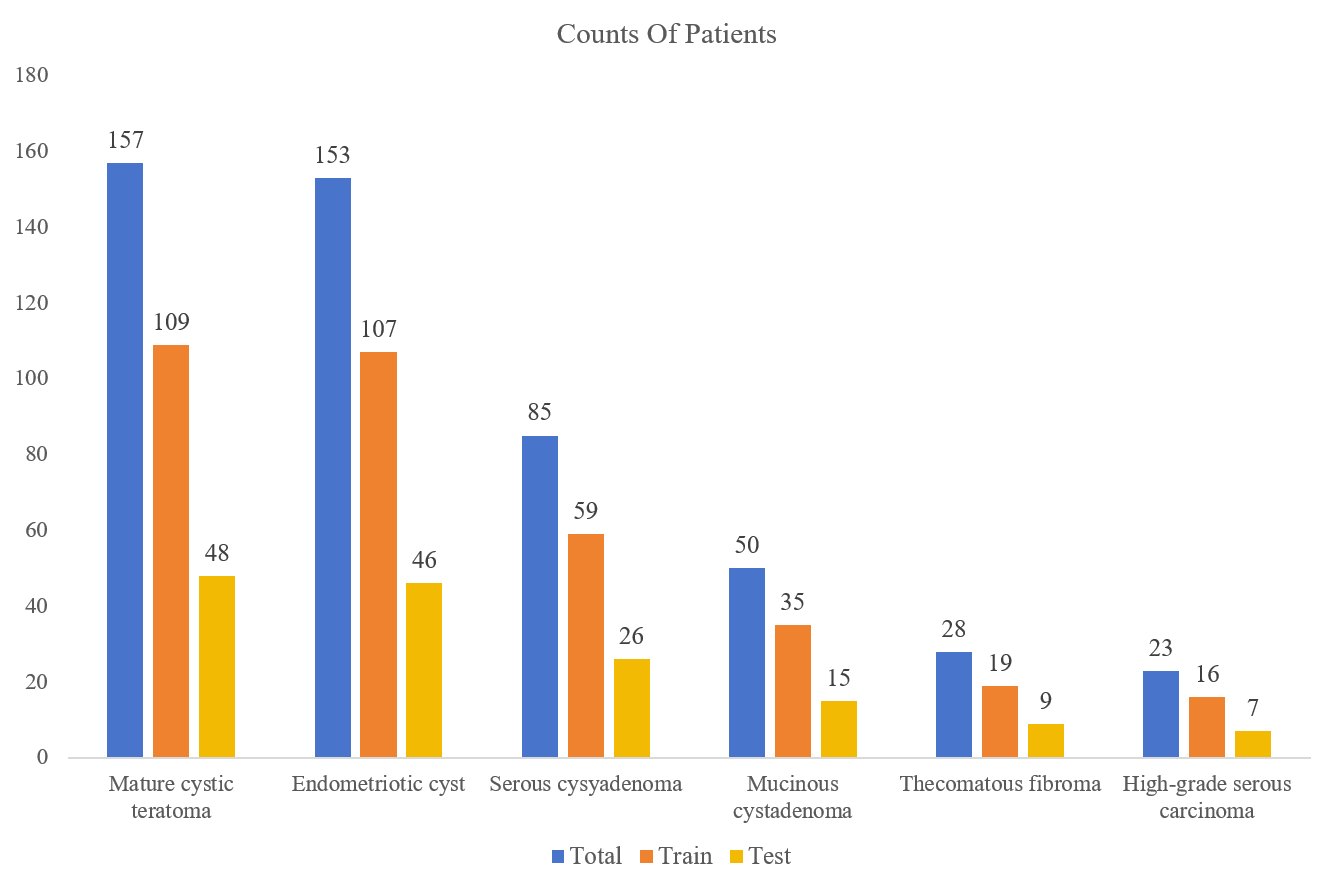}
\caption{The number of patients containing in each category. The horizontal axis of the table corresponds to category and the left vertical axis shows the count.}
\label{patient}
\end{figure}
\subsection{Data Analysis}
In ViTaL image dataset, all images have been analyzed by experts using specialized software, so some symbols like “lines” , “hands” , “characters” or “arrows” are marked by them on images. The basic typical modality data of the dataset are illustrated in the figure\ref{example}. Specifically, the image subset contains 2216 traditional 2D ultrasound images, shown as Fig.\ref{example}. For every images, only one type of tumor appears. The tabular modality data primarily consists of information collected from patients during relevant examinations, including the patient's age, BMI, abdominal pain, abdominal bloating, the values of five tumor markers, and the maximum diameter of the tumors. This tabular data is meticulously organized to ensure that each piece of information is accurately recorded and easily accessible for analysis. The textual data primarily comprises medical reports generated based on ultrasound images, focusing on the description of certain aspects of ovarian tumors. These reports are integral to the dataset as they provide detailed and qualitative insights that complement the quantitative data from the tabular and image modalities. The integration of these detailed metrics with other modalities such as images and text allows for a more nuanced and robust approach to patient diagnosis and treatment planning. By leveraging this multi-modal dataset, researchers and healthcare professionals can gain deeper insights into patient health, leading to improved outcomes in clinical practice. 
As shown in Figures 3 and 4, the dataset is primarily composed of visual, tabular and linguistic data from 496 patients. In total, there are six types of ovarian tumors in this dataset, namely Mature cystic teratoma, Endometriotic cyst, Serous cystadenoma, Mucinous cystadenoma, Thecomatous fibroma, and High-grade serous carcinoma. The images in the dataset exhibit a variety of proportions, with the width ranging from 302 to 1135 pixels and the height ranging from 226 to 794 pixels. To standardize the image dimensions for training purposes, each image undergoes a preprocessing step where it is first randomly resized and then cropped to a uniform size of 224×224 pixels. This process ensures that the model can effectively process the images regardless of their original size, thereby maintaining consistency and improving the efficiency of the training process.
\begin{figure}[t]
    \centering
    \includegraphics[width=\linewidth]{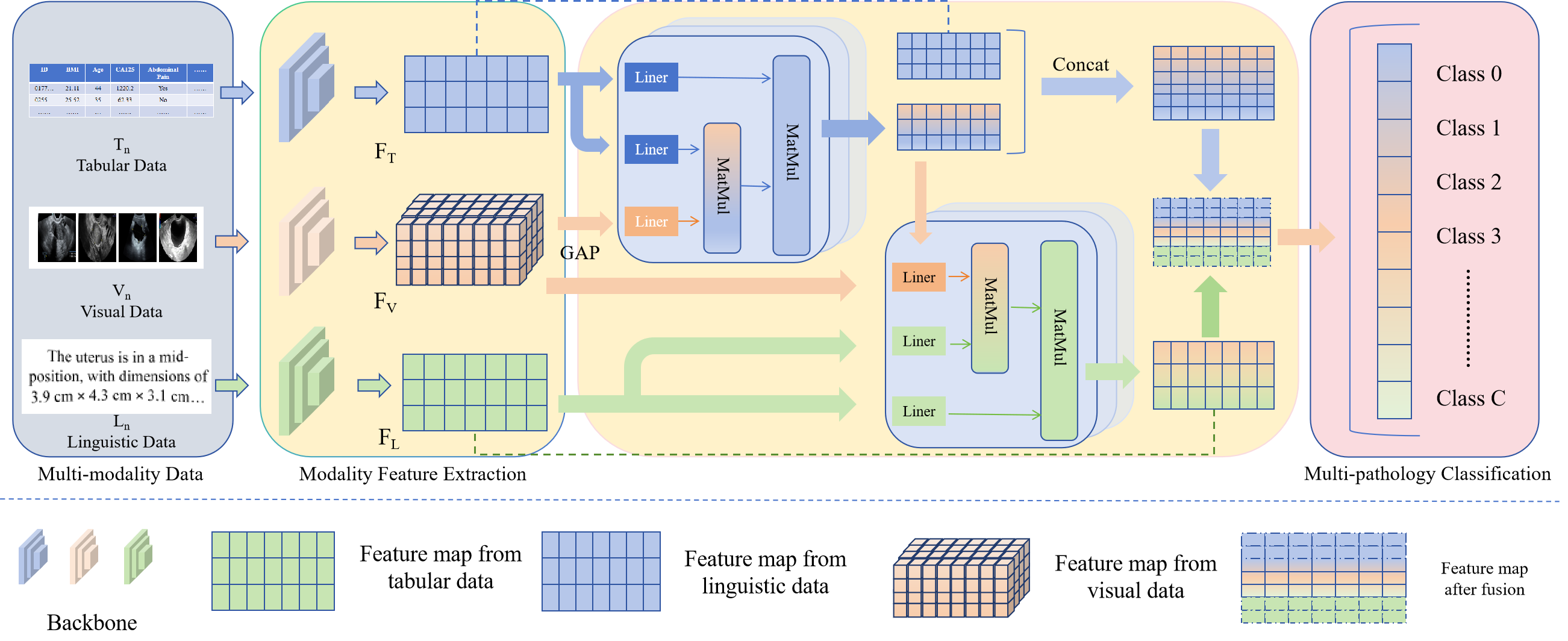}
    \caption{Overall structure of the proposed ViTaL-Net. It includes three components: feature extraction for each modality, fusion of features from different modalities, and the final multi-pathology classification.}
    \label{flow}
\end{figure}
\section{Method}
In this section, we first provide a brief overview of the proposed network and then detail each component in the following subsections. 
The flowchart of our method is depicted in Fig.\ref{flow} and includes three steps. First, every single-modality feature extraction network is employed to extract features from different modality data. Then, Triplet Hierarchical Offset Attention Mechanism (THOAM) is used to fuse the extracted features from different modalities, aiming to minimize the loss incurred during the information fusion process. Finally, the synthesized feature map is classified to predict ovarian tumors.
\subsection{Preliminary}
The task of ovarian tumor classification is, in principle, a multi-pathological classification problem. Let \(\mathcal{D} = \{(\mathbf{V}_n^i, \mathbf{T}_n, \mathbf{L}_n), y_n\}_{n=1}^N\) denote the constructed multi-modal dataset $\mathcal{D}$, where \(N\) is the number of samples, \(y_n \in \{1, 2, \cdots, C\}\) represents the label of each samples, $\mathbf{V}_n^i, \mathbf{T}_n, \mathbf{L}_n$ represents $i$-th slice visual, tabular and linguistic information data for the $n$-th sample and \(C\) is the total number of classes.
\par Given a function space \( \mathcal{F} \) for a Neural Network used in multi-class classification, the goal of training this network is to find a classifier function \( f \in \mathcal{F} \) that can effectively map the input multi-modality representations of ovarian tumor data to appropriate predictions or scores. The role of the classifier is to assign each input a label or score indicating the likelihood of belonging to a specific ovarian tumor category. This multi-pathological multi-modality ovarian tumor classification can be defined by the following minimization problem:\begin{equation}
    \underset{f \in \mathcal{F}}{\operatorname{argmin}} \frac{1}{N} \sum_{n = 1}^{N} \mathcal{L}(f(D_n), y_n)
    \label{opt}
\end{equation}
where $D_n$ and $y_n$ represent the visual, tabular and linguistic input data for the n-th training sample and the corresponding label. $\mathcal{L}$ is the loss function for each sample. Each term in the sum represents the loss for a specific training sample, and the objective is to minimize the average loss over the entire training set.
\subsection{Single-modality feature extraction}
For different modalities, we select their respective backbones to extract features. We use the MobileNet model to extract features from the image modality data. MobileNet is a lightweight and efficient convolutional neural network that is well-suited for image feature extraction, providing a good balance between computational efficiency and accuracy. It is designed to quickly and accurately process visual information, extracting features that are essential for downstream tasks such as image classification and object detection. For the tabular modality data, we employ the TabNet model. TabNet is specifically designed to handle tabular data and can learn meaningful representations by selectively focusing on important features. As for the linguistic modality data, we utilize BERT to encode the text in the text modality, thereby obtaining its feature map. BERT is a powerful pre-trained language model that can effectively capture the semantic information and contextual relationships within the text, resulting in a high-quality feature representation. In summary, by selecting MobileNet, TabNet and BERT as the backbones for the visual, tabular and linguistic modalities respectively, we ensure that each modality unique characteristics are effectively captured and represented.
\subsection{Triplet Hierarchical Offset Attention Mechanism}
Triplet Hierarchical Offset Attention Mechanism is a model that extends the traditional cross-attention mechanism and is used to handle the interaction relationships among three input modalities. This mechanism not only captures the individual features of each modality but also effectively integrates and aligns them in a hierarchical manner. By introducing offset attention, it further enhances the ability to focus on specific regions or elements within each modality, thereby improving the overall performance and accuracy of multi-modal data processing. As shown in the Fig.~\ref{flow}, the data from each modality is processed through its respective Backbone to obtain the corresponding feature maps $\left ( F_V,F_T,F_L \right ) $. The feature maps of the image $F_V\in R^{N\times C\times H\times W}$ are integrated in the spatial dimension of the feature maps by global average pooling:
\begin{equation}
\text{GAP}(F_V) = \frac{1}{H \times W} \sum_{i = 1}^{H} \sum_{j = 1}^{W} F_{i,j}
\label{gap}
\end{equation}where \(H\) and \(W\) represent the height and width of the feature map, respectively. After pooling, the feature maps from the three modalities \((F'_V, F_T, F_L) \in \mathbb{R}^{N \times C}\) start to be processed through attention mechanism. As shown in Fig.~\ref{flow}, given the feature maps of the image and tabular modalities $F'_V,F_T$,  the matrices of query Q, key K and values V are first calculated through three different linear projections.The main calculation process is as follow:
\begin{equation}
\begin{aligned}
    Q_1 = Linear\left ( F_V \right ) =F_VW_Q\\
    K_1 = Linear\left ( F_T \right ) =F_TW_K\\
    V_1 = Linear\left ( F_T \right ) =F_TW_V\\
\end{aligned}
\end{equation}where $W_Q,W_K,W_V$ are the corresponding weight matrices of linear projections. Subsequently, the key (K) is transposed and multiplied with the query (Q) to generate the raw scores. These raw scores represent the initial similarity measurements between the query and the key vectors. To convert these raw scores into a probability distribution that reflects the relative importance of each key, we apply the softmax function. The softmax function normalizes the raw scores, ensuring that they sum up to 1 and can be interpreted as probabilities. This step is critical because it allows the model to focus on the most relevant parts of the input data by assigning higher weights to the keys that are more similar to the query. The scores represent the weighted importance of each key relative to the query, main calculation process is as follow:\begin{equation}
    Score= Softmax\left ( \frac{Q_1K_1^T}{\sqrt{d_m}}  \right ) 
    \label{qk1}
\end{equation}where $\sqrt{d_m}$ is the s a scaling factor. Following the computation of the scores using the softmax function, we multiply these scores by the value (V) matrix to obtain a weighted sum of the values. This operation allows us to integrate the information from the values, where each value is weighted by its corresponding score, reflecting its relevance to the query. The result of this weighted sum is then passed through a linear transformation, which can be thought of as a final layer that maps the output of the attention mechanism to the desired output space.\begin{equation}
    F_1 = Linear\left ( ScoreV_1 \right )
    \label{sv}
\end{equation}where $Score, V$ represent the previously computed scores and the value vectors, respectively. The features of the visual modality information and the tabular modality information have been fused. The feature map at this moment is referred to as $F_1$. Then, $F_1$ is further interacted with feature from the linguistic modality. This interaction aims to fuse the visual and linguistic information, allowing the model to better understand the context and semantics by leveraging the complementary features from both modalities.\begin{equation}
\begin{aligned}
    Q_2 = Linear\left ( F_1 \right ) =F_1W_Q\\
    K_2 = Linear\left ( F_L \right ) =F_LW_K\\
    V_2 = Linear\left ( F_L \right ) =F_LW_V\\
\end{aligned}
\end{equation}where $W_Q, W_K , W_V$ are the new corresponding weight matrices of linear projections, which are different from the previous ones. Subsequently, the key ($K_2$) is transposed and then multiplied with the query ($Q_2$), thereby generating the raw scores. These raw scores act as the initial metrics of similarity between the query and the key vectors. Afterward, these raw scores are passed through a softmax function to obtain the final scores. These final scores are subsequently multiplied with the value (V) through matrix multiplication, resulting in the final feature map. The final output, after undergoing the series of transformations and interactions, is concatenated with the initial visual feature map. This concatenation step is crucial as it integrates the refined, multi-modal information back into the original visual context, thereby enriching the representation with both the enhanced semantic understanding and the spatial details from the visual feature map. 
\begin{equation}
\scalebox{0.96}{
    $Output = Concact[Linear(Softmax\left ( \frac{Q_2K_2^T}{\sqrt{d_m}}  \right ) V2),F'_V]$
    }
\end{equation}where $\sqrt{d_m}$ is the s a scaling factor, $F_V$ is the feature map of visual modality data. The final $Output\in R^{N\times C}$ is achieved by fusing the feature maps from three modalities. This fusion process integrates the unique strengths and information from each modality, resulting in a comprehensive and enriched representation that captures the nuances and details from all three sources.
\subsection{Decoder and Loss Function}
We employ a linear decoder to decode the feature maps achieved by fusing the feature maps from three modalities.\begin{equation}
y_i = \sum_{j = 1}^{d} W_{i,j} \cdot h_j + b_i
\end{equation}where \( y_i \) is the predicted value for each category, \( W_{ij} \) is the matrix of the linear classification decoder, and \( b \) is the bias. $h_j$ represents the values of each column in the input feature map.
During the training stage, we employ a combination of the cross-entropy loss as the loss function, defined as Equ.\ref{loss} :\begin{equation}
    \mathcal{L} = -\frac{1}{N} \sum_{n}^N [y_n \cdot \log(p_n) + (1 - y_n) \cdot \log(1 - p_n)]
    \label{loss}
\end{equation}
where $p_n$ refer to the probability that the prediction of the $n$-th sample and $y_n$ is the true label for the $n$-th sample.
\section{Experiments}
In this section, we first introduce the data processing methods, implementation details, and evaluation metrics. We then compare ViTaL-Net with several state-of-the-art classification methods and provide both qualitative and quantitative experimental results. Finally, we conduct ablation studies to verify the effectiveness of each component in our method.
\subsection{Data Processing}
We performed the following processing on proposed ViTaL for network training and testing. Before training, input ultrasound images are randomly resized and cropped to 224 × 224. In the experiments, we applied data augmentations of horizontal flips, vertical flips, and random rotations to reduce overfitting for visual data. For tabular and linguistic data, we do not perform data augmentation.
\par For tabular data, we extracted 10-dimensional data from the statistical tabular data, which includes age, BMI, abdominal pain, abdominal bloating, the values of five tumor markers, and the maximum diameter. For the tabular data of age, BMI and the maximum diameter, which have certain ranges, we perform normalization using the maximum and minimum values, shown as Equ.\ref{maxmin} :
\begin{equation}
    x = \frac{x-x_{min}}{x_{max}-x_{min}} 
    \label{maxmin}
\end{equation}
where $x_{min}$ and $x_{max}$ represent the maximum and minimum values of that data in the entire sample set.
\par For abdominal pain, abdominal bloating these types of data that only have "yes" or "no" categories, we apply 0-1 encoding, setting the data for abdominal pain and abdominal bloating to 1 if present and 0 if absent. The remaining five types of tumor marker data are CA-125, CEA, CA199, AFP, and CA153. For some malignant tumors, these five types of tumor marker data may exhibit unusual values. For the values of CEA and AFP, we apply Gaussian normalization to mitigate the impact of outliers on the entire dataset, shown as Equ.\ref{gas} :
\begin{equation}
    x = \frac{x-\mu}{\sigma } 
    \label{gas}
\end{equation}
\begin{table}[t]
\caption{Implementation details of all models. All training processes adopt a StepLR learning rate decay strategy.}
\begin{center}
\scalebox{0.9}
{    \begin{tabular}{ccccc}
    \hline
    Methods      & Optimizer & Initial-LR & StepLR Decay & Epoch \\ \hline
    ResNet~\cite{he2016deep}       & SGD       & 0.001      & 30,60,90     & 100   \\ \hline
    Densenet~\cite{huang2017densely}     & SGD       & 0.001      & 30,60,90     & 100   \\ \hline
    SeResNet~\cite{hu2018squeeze}     & SGD       & 0.001      & 30,60,90     & 100   \\ \hline
    Shufflenet~\cite{zhang2018shufflenet}   & SGD       & 0.001      & 30,60,90     & 100   \\ \hline
    Moblienet~\cite{howard2017mobilenets}    & SGD       & 0.001      & 30,60,90     & 100   \\ \hline
    ViT~\cite{dosovitskiy2020image}          & AdamW     & 0.001      & 30,60,90     & 100   \\ \hline
    Efficientnetv2~\cite{tan2021efficientnetv2} & AdamW     & 0.001      & 30,60,90     & 100   \\ \hline
    \label{9}
    \end{tabular}
}
\end{center}
\end{table}
where $\mu$ and $\sigma$ represent the mean and standard deviation of that data in the entire sample set.
\par The remaining three marker values can exhibit extremely high values in some samples. For example, CA125 levels may reach as high as 9000 in some samples, while in others they may only be as low as 4. Therefore, we employ a more robust normalization method, Robust Scaling, which can significantly reduce the interference of outliers on the entire dataset, shown as Equ.\ref{rob} :
\begin{equation}
    x = \frac{x-x_{median}}{x_{Q_3}-x_{Q_1}} 
    \label{rob}
\end{equation}
where $x_{Q_3}$ and $x_{Q_1}$ represent the upper quartile and lower quartile of that data in the entire sample set, and $x_{median}$ is the median of that data in the entire sample set.
\subsection{Implementation Details and Evaluation Metrics}
\begin{table}[t]
\caption{The classification performance of different models, including ViTaL-Net, on the dataset was evaluated, with the AUC values and accuracy rates for various tumor pathologies being reported. T.F, E.C, M.C.T, S.C, H.G.S.O.C, and M.C represent Thecoma fibroma, Endometriotic cyst, Mature cystic teratoma, Serous cystadenoma, High-grade serous carcinoma, and Mucinous cystadenoma, respectively. The best values are highlighted in bold.}
\begin{center}
\scalebox{0.8}{
    \begin{tabular}{c|cccccc|c}
    \hline
                            & \multicolumn{6}{c|}{AUC(Area Under the Curve)}                                                                                               &                                \\ \cline{2-7}
    \multirow{-2}{*}{Model} & T.F & E.C & M.C.T      & S.C & H.G.S.C & M.C & \multirow{-2}{*}{Accuracy(\%)} \\ \hline
    Densenet121~\cite{huang2017densely}             & 0.87            & 0.96               & {\color[HTML]{000000} 0.91} & 0.91               & 0.83                        & 0.76                 & 71.02                          \\
    Resnet34~\cite{he2016deep}                & 0.86            & 0.96               & 0.92                        & 0.92               & 0.77                        & 0.85                 & 71.02                          \\
    Resnet50~\cite{he2016deep}                & 0.86            & 0.94               & 0.87                        & 0.90               & 0.75                        & 0.75                 & 66.33                          \\
    SeResnet50~\cite{hu2018squeeze}              & 0.78            & 0.95               & 0.89                        & 0.92               & 0.82                        & 0.87                 & 70.69                          \\
    ShufflenetV1~\cite{zhang2018shufflenet}            & 0.80            & 0.96               & 0.88                        & 0.88               & 0.81                        & 0.85                 & 67.67                          \\
    \rowcolor[HTML]{EFEFEF} 
    Moblienets2~\cite{howard2017mobilenets}             & 0.71            & 0.98               & 0.94                        & 0.91               & 0.84                        & 0.85                 & 74.54                          \\
    ViT~\cite{dosovitskiy2020image}                     & 0.91            & 0.96               & 0.90                        & 0.92               & 0.81                        & 0.83                 & 68.88                          \\
    Efficientnetv2~\cite{tan2021efficientnetv2}          & 0.80            & 0.95               & 0.87                        & 0.92               & 0.85                        & 0.81                 & 69.07                          \\ \hline
    CBAM~\cite{woo2018cbam}                    & 0.88            & \textbf{0.98}      & 0.94                        & 0.94               & 0.88                        & 0.89                 & 76.04                          \\
    ITCM~\cite{hu2025multi}                   & 0.82            & \textbf{0.98}      & 0.91                        & \textbf{0.95}      & 0.86                        & 0.89                 & 73.71                          \\
    TFA-LT~\cite{li2024text}                   & 0.89            & \textbf{0.98}      & 0.96                        & \textbf{0.95}      & 0.88                        & \textbf{0.92}        & 76.72                          \\
    \rowcolor[HTML]{C0C0C0} 
    \textbf{THOAM(Ours)}    & \textbf{0.98}   & \textbf{0.98}      & \textbf{0.99}               & 0.93               & \textbf{0.98}               & 0.82                 & \textbf{83.08}                 \\ \hline
    \end{tabular}
}
\label{aucacc}
\end{center}
\end{table}
\subsubsection{Implementation Details}
As shown in Tab.\ref{9}, we utilized the SGD and AdamW optimizer to update our models with a weight decay of 1e-5 for training. We use Cross Entropy Loss to train all models. The total number of epochs was set to 100 with an initial learning rate of 1e-3 and batchsize of 32. To ensure fairness, we employed the same learning rate decay strategy for training all models, with the learning rate decaying at a rate of 0.3 when the training epoch reaches 30, 60 and 90. All training were based on Pytorch and MMpretrain. We use a Nvidia RTX 4080 Super GPU for all experiments. Tab.\ref{9} reports the detailed model settings.
\subsubsection{Evaluation Metrics}
We evaluated our experiments based on accuracy (ACC), area under the receiver operating characteristic curve (AUC), specificity (SPE) and sensitivity (SEN). All of these metrics have values ranging from 0 to 1, and a higher value indicates better model performance.

\subsection{Compare Methods}
We first conducte image unimodal recognition and classification on eight basic models, shown as Tab.~\ref{aucacc} and Tab.~\ref{spesen}. Then, we selected the best baseline model (Moblienets2~\cite{howard2017mobilenets}) for the visual network to conduct multimodal classification experiments. Due to the limited research on the fusion of visual, tabular and linguistic modalities, we selected some fusion methods based on image and text (TFA-LT~\cite{li2024text}), image and tabular data (ITCM~\cite{hu2025multi}) as well as CBAM~\cite{woo2018cbam} for comparative experiments.
\subsection{Experiment Results}
\begin{table}[]
\caption{The classification performance of different models, including ViTaL-Net, on the dataset was evaluated, with the specificity and sensitivity for various tumor pathologies being reported. T.F, E.C, M.C.T, S.C, H.G.S.O.C, and M.C represent Thecoma fibroma, Endometriotic cyst, Mature cystic teratoma, Serous cystadenoma, High-grade serous carcinoma, and Mucinous cystadenoma, respectively. All values in the table are expressed as percentages (\%). The best values are highlighted in bold.}
\begin{center}
\scalebox{0.62}{
    \begin{tabular}{c|clccccccccccc}
    \hline
                            & \multicolumn{3}{c}{T.F}                                              & \multicolumn{2}{c}{E.C}        & \multicolumn{2}{c}{M.C.T} & \multicolumn{2}{c}{S.C} & \multicolumn{2}{c}{H.G.S.C} & \multicolumn{2}{c}{M.C} \\ \cline{2-14} 
    \multirow{-2}{*}{Model} & \multicolumn{2}{c}{SEN}                                & SPE             & SEN                      & SPE        & SEN              & SPE            & SEN            & SPE           & SEN                & SPE               & SEN            & SPE            \\ \hline
    Densenet121~\cite{huang2017densely}             & \multicolumn{2}{c}{26.67}                                  & 98.94               & 85.47                        & 93.18          & 77.78                & 87.72               & 73.77              & 89.68             & 29.41                  & 95.74                  & 36.59               & 97.12              \\
    Resnet34~\cite{he2016deep}                & \multicolumn{2}{c}{36.67}                                  & 98.77               & {\color[HTML]{000000} 81.98} & \textbf{97.88}          & 86.36                & 80.21               & 63.11              & 91.37             & 5.88                   & 99.82                  & 53.66               & 93.53              \\
    Resnet50~\cite{he2016deep}                & \multicolumn{2}{c}{26.67}                                  & 98.59               & 81.98                        & 92.47          & 68.69                & 89.22               & 70.49              & 87.79             & 26.47                  & 97.69                  & 39.02               & 91.55              \\
    SeResnet50~\cite{hu2018squeeze}              & \multicolumn{2}{c}{16.67}                                  & \textbf{99.82}      & 84.88                        & 97.18          & 79.29                & 80.95               & 81.15              & 85.26             & 0                      & \textbf{99.99}         & 36.59               & 97.12              \\
    ShufflenetV1~\cite{zhang2018shufflenet}            & \multicolumn{2}{c}{16.67}                                  & 98.41               & 83.14                        & 93.18          & 81.82                & 79.95               & 62.30              & 92.01             & 5.88                   & 99.11                  & 39.02               & 94.24              \\
    \rowcolor[HTML]{EFEFEF} 
    Moblienets2~\cite{howard2017mobilenets}             & \multicolumn{2}{c}{\cellcolor[HTML]{EFEFEF}40.00}          & 97.71               & 90.70                        & 94.82          & 83.84                & 88.22               & 72.13              & 90.53             & 26.47                  & 99.29                  & 34.15               & 96.22              \\
    
    ViT~\cite{dosovitskiy2020image}                     & \multicolumn{2}{c}{30.00}                                  & 98.24               & 76.16                        & 96.24          & 76.77                & 85.71               & 86.07              & 82.11             & 5.88                   & 99.47                  & 29.27               & 97.30              \\
    Efficientnetv2~\cite{tan2021efficientnetv2}          & \multicolumn{2}{c}{26.67}                                  & 98.77               & 86.05                        & 92.71          & 81.31                & 77.94               & 68.85              & 91.37             & 2.94                   & \textbf{99.99}         & 24.39               & 96.76              \\ \hline
    CBAM~\cite{woo2018cbam}                    & \multicolumn{2}{c}{30.00}                                  & 98.94               & 90.70                        & 96.24          & 85.86                & 86.72               & 78.69              & 90.74             & 5.88                   & 99.47                  & 51.22               & 96.22              \\
    ITCM~\cite{hu2025multi}                    & \multicolumn{2}{c}{26.67}                                  & 97.71               & 82.56                        & 96.24          & 84.34                & 83.96               & 75.41              & 93.47             & 14.71                  & 99.47                  & \textbf{63.41}      & 94.60              \\
    TFA-LT~\cite{li2024text}                  & \multicolumn{2}{c}{13.33}                                  & 99.65               & 90.12                        & 95.76          & 89.90                & 81.20               & 81.15              & \textbf{93.68}    & 2.94                   & 99.82                  & 51.22               & \textbf{97.66}     \\
    \rowcolor[HTML]{C0C0C0} 
    \textbf{THOAM(Ours)}         & \multicolumn{2}{c}{\cellcolor[HTML]{C0C0C0}\textbf{53.33}} & 97.71               & \textbf{93.02}               & 96.47 & \textbf{91.41}       & \textbf{98.25}      & \textbf{79.51}     & 90.53             & \textbf{82.35}         & \textbf{99.99}         & 34.15               & 96.22              \\ \hline
    \end{tabular}
}
\end{center}
\label{spesen}
\end{table}
In Tab.~\ref{aucacc}, we present the classification performance of the proposed framework and comparison methods on six ovarian tumor-related classification tasks, namely the diagnosis of ovarian tumor lesions with six classifications: Mature cystic teratoma, Endometriotic cyst, Serous cystadenoma, Mucinous cystadenoma, Thecomatous fibroma, and High-grade serous carcinoma. As can be seen, in the majority of the six ovarian tumor lesion classification tasks, the proposed ViTaL-net achieves the best results among all methods. Specifically, in the ovarian tumor diagnosis task on the ViTaL dataset, the proposed framework obtains a high performance of 14.5\% higher than the best competitor in terms of Acc. After incorporating tabular and textual data, the model is able to identify tumor data from more perspectives, thereby further enhancing its accuracy. For instance, in the category of high-grade serous carcinoma, tabular data can identify this pathological type based on symptoms such as abdominal pain or bloating. 
As shown in the Tab.\ref{spesen}, ViTaL-Net also exhibits excellent performance, with high specificity and sensitivity. In absolute terms, while ViTaL-Net may not always achieve the highest performance across every single metric, its comprehensive performance is quite remarkable. When we look at other models, we often find a trade-off between specificity and sensitivity. Some models may exhibit high specificity but suffer from extremely low sensitivity, meaning they are good at identifying non-cancerous cases but miss a significant number of actual cancer cases. On the other hand, some models may have high sensitivity, effectively detecting a large proportion of cancer cases, but their low specificity leads to a high number of false positives, misidentifying many non-cancerous cases as cancerous.
In the clinical context of identifying ovarian tumors, the implications of these imbalances are profound. Models with poor sensitivity would result in a high rate of false negatives, meaning many patients with ovarian tumors would be incorrectly told they are cancer-free, potentially delaying crucial treatment. Conversely, models with poor specificity would produce a high rate of false positives, leading to unnecessary anxiety, further invasive testing, and possibly even overtreatment for patients who do not actually have cancer. Neither scenario is acceptable in a clinical setting, where accurate diagnosis is paramount for effective patient care and management.
\subsection{Visualization Analysis}
\begin{figure}[t]
    \centering
    \includegraphics[width=\linewidth]{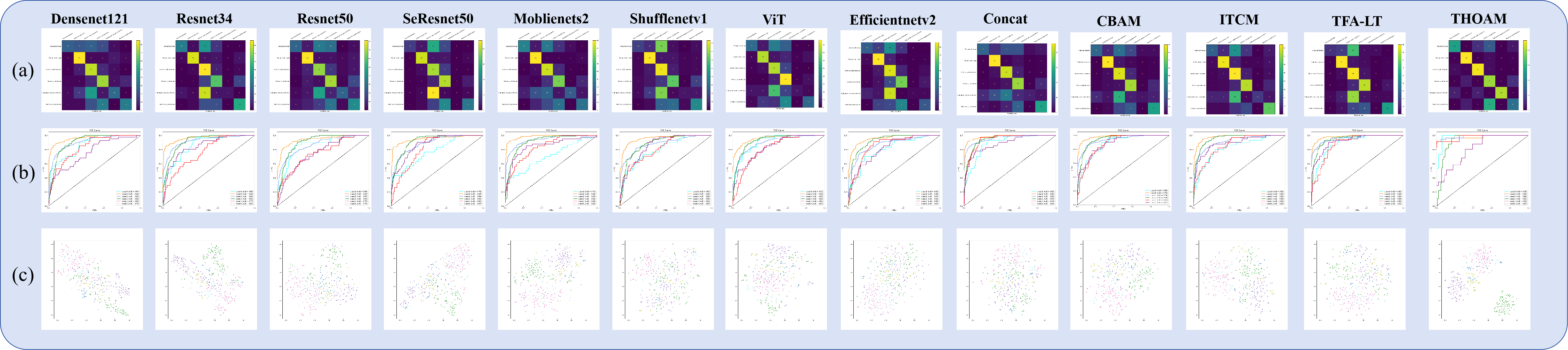}
    \caption{Visualization of classification performance on the ViTaL dataset using various classification models and different modality feature fusion methods. Sub-figures (a), (b) and (c) represent confusion matrices, ROC curves, and t-SNE plots, respectively.}
    \label{vis}
\end{figure}
\par Fig.~\ref{vis} presents the performance evaluation results of various neural network models on the ovarian tumor identification task within this dataset, showcasing the outcomes through three aspects: confusion matrices, ROC curves, and t-SNE visualizations. Due to the limited number of samples in some categories, certain classes may have high recognition rates, but there may not be significant changes in the confusion matrix. Therefore, we have preprocessed the values in the confusion matrix by normalizing each row, dividing the values in each row by the sum of that row. In this way, even if some classes have a small number of samples, their true classification performance on those classes can be better demonstrated. In Fig.~\ref{vis}, we can clearly observe that the performance of ViTaL-Net has been significantly enhanced and improved compared to other directional models. Fig.~\ref{vis}(a) illustrates the classification performance of the confusion matrices for different models, where the confusion matrix corresponding to ViTaL-Net has the most prominent main diagonal color. This indicates that ViTaL-Net has achieved better classification performance across various categories. 
As shown in the Fig.~\ref{vis} (b), in the ROC curves corresponding to each model, ViTaL-Net also exhibits significant improvement. For many pathological categories, its AUC values are increasingly approaching 1, indicating that it performs better in distinguishing between positive and negative samples. In the t-SNE visualization, different colors of points represent different categories. For ViTaL-Net, points of the same category are clustered together, while points of different categories are well-separated. This indicates that the model is capable of effectively extracting features and classifying the data. In contrast, the visualizations for other models show points mixed together, which suggests that these models have difficulty in distinguishing between different categories.
\begin{figure}[t]
    \centering
    \includegraphics[width=\linewidth]{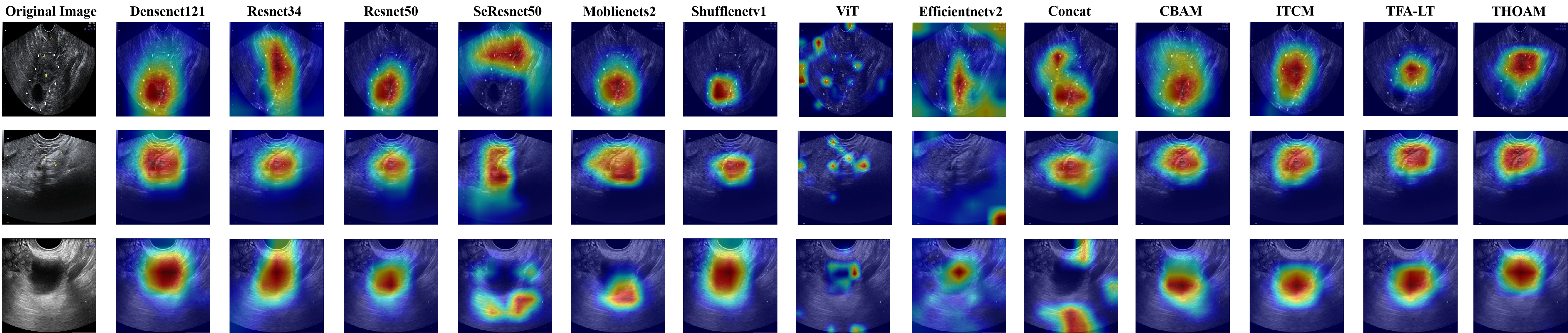}
    \caption{Visualization of heatmaps for different pathological images using various classification models and different modality feature fusion methods.}
    \label{cam}
\end{figure}
\par As illustrated in the Fig.~\ref{cam}, the activation mapping for each model is presented. The colors in these maps signify the degree of attention that the model pays to different regions within the image. Generally, warm colors such as red are indicative of areas that the model focuses on intensely. This heightened attention suggests that these particular regions play a crucial role in shaping the model's classification decisions. Conversely, cool colors like blue highlight areas that garner relatively less attention from the model. We can clearly observe that the incorporation of textual and tabular modalities aids the visual modality in generating feature maps that focus more intently on regions of tumors.

\subsection{Ablation Study}
\begin{table}[]
\begin{center}
\caption{Performance (ACC and AUC) of Different Combinations of Visual, Tabular and Linguistic Modalities. The AUC values in the table are the averages for each pathological category.}
\scalebox{0.85}{
    \begin{tabular}{ccc|cc}
    \hline
    \multicolumn{3}{c|}{Modality} &                       &                       \\ \cline{1-3}
    Visual & Tabular & Linguistic & \multirow{-2}{*}{ACC(\%)} & \multirow{-2}{*}{AUC} \\ \hline
    $\checkmark$       & $\times$       & $\times $         & 74.54 &0.87  \\
    $\times$      & $\checkmark$       & $\times$          &59.46  &0.84  \\
    $\times$      & $\times$       & $\checkmark$          &69.01  &0.85  \\
    $\checkmark$       & $\checkmark$      & $\times$          &78.39  &0.94 \\
    $\checkmark $      & $\times$      & $\checkmark$          &78.89  &0.92 \\
    $\times$     & $\checkmark$        & $\checkmark$          &80.40  &0.92 \\
    \rowcolor[HTML]{C0C0C0} 
    $\checkmark$       & $\checkmark$       & $\checkmark$         &\textbf{85.59}  &\textbf{0.95}  \\ \hline
    \end{tabular}
}
\label{ablation}
\end{center}
\end{table}
To verify the effectiveness of feature extraction module for each modality, we conducted ablation studies to better understand the impacts of each component. We give the qualitative and quantitative classification results and analyze each module. We completed the experiments based on the removal of the extraction of different modality data. After sequentially removing different modalities of input, we measure the final prediction accuracy of the model and the average AUC for each category. The specific ablation experiments are shown in the Tab.~\ref{ablation}. It can be observed that the performance of individual modalities used as single-modality inputs for separate pathological recognition is not very satisfactory. After incorporating an additional modality, the performance of each modality sees a certain degree of improvement. When all three modalities are integrated, the model achieves its optimal performance. Each modality possesses information from different perspectives, which can be provided to the model to facilitate the identification of ovarian tumors from various angles. For instance, from the perspective of visual modality, model can obtain information regarding the shape and size of ovarian tumors, as well as discern whether the surface of the ovarian tumor is smooth. These characteristics may vary depending on the pathological type of the ovarian tumor. However, the information from the tabular data can reveal whether a patient's examination markers are normal and if the patient has any abdominal pain. For example, patients with high-grade serous carcinoma are difficult to distinguish at the image level, but the model can identify this category based on certain markers and abdominal pain in the tabular data. The linguistic modality data mainly provides information on the relative position, size, and shape of ovarian tumors, which also offers certain perspectives for the model to identify ovarian tumors. 
\par In summary, the integration of multiple modalities significantly enhances the model's ability to accurately classify and identify ovarian tumors. The complementary nature of the information provided by each modality allows the model to overcome the limitations of single-modality inputs, particularly in distinguishing categories with fewer samples and lower recognition rates. This comprehensive approach not only improves overall classification performance but also provides a more robust and reliable framework for ovarian tumor identification.
\section{Discussion}
\subsection{Concatenation vs Attention Fusion}
In multimodal feature fusion, there are also some relatively simple and commonly used methods, such as concatenation. We directly concatenate the feature data of the image after average pooling with the feature data of the table and text to form a feature map containing information from three modalities and then proceed with classification. As shown in the Tab~\ref{dis}, we applied this method on the ViTaL dataset to compare with our proposed THOAM in terms of experimental effects. It can be observed that the classification performance of the model with direct concatenation is not satisfactory. Even though simple concatenation does not perform poorly in many applications, on the ViTaL dataset, the large differences in feature representations among different modalities may cause significant loss in the fusion of information from different modalities when using simple concatenation, which ultimately leads to poor classification results. Attention-based mechanism we proposed can effectively enhance the correlation and complementarity between different modalities. By focusing on the most relevant and informative features from each modality, this mechanism enables more efficient and meaningful integration of multi-modality data.
\begin{table}[]
\caption{Comparison of model classification performance before and after incorporating the “Other” category.}
\label{dis}
\centering
\scalebox{0.85}{
    \begin{tabular}{c|ccc|cc}
    \hline
    \multicolumn{1}{c|}{\multirow{2}{*}{Dataset (ViTaL)}} & \multicolumn{3}{c|}{Modality}                                                              & \multirow{2}{*}{ACC(\%)} & \multirow{2}{*}{AUC} \\ \cline{2-4}
    \multicolumn{1}{c|}{}                         & \multicolumn{1}{l}{Visual} & \multicolumn{1}{l}{Tabular} & \multicolumn{1}{l|}{Linguistic} &                          &                      \\ \hline
    Concatenation                             & $\checkmark$               & $\checkmark$                & $\checkmark$                    & 75.88          & 0.93        \\
    Attention Fusion                           & $\checkmark$               & $\checkmark$                & $\checkmark$                   & \textbf{85.59}           & \textbf{0.95}               \\ \hline
    Including “Other"                             & $\checkmark$               & $\checkmark$                & $\checkmark$                    & 70.39                    & 0.86                 \\ 
    Excluding “Other"                             & $\checkmark$               & $\checkmark$                & $\checkmark$                    & \textbf{85.59}           & \textbf{0.95}        \\ \hline
    \end{tabular}
}
\end{table}
\subsection{Excluding “Other” vs Including “Other”}
The ViTaL dataset comprises a diverse range of pathological categories, specifically six well-defined classes, each representing distinct pathological conditions. Additionally, there is an seven category labeled as “Other” which encompasses a variety of less common pathological samples. This “Other” category is particularly heterogeneous, including a wide range of rare and unique pathological cases that do not fit neatly into the more established categories.
The inclusion of this “Other” category in the model training presents significant challenges. Due to the high variability and lack of common features among the samples in this category, incorporating it into the model can lead to a substantial decrease in overall model performance, shown as Tab.\ref{dis}. The diverse nature of the “Other” category introduces noise and complexity, making it difficult for the model to learn consistent and meaningful patterns. This, in turn, can negatively impact the model's ability to accurately classify the more well-defined categories. Moreover, the limited number of samples in the “Other” category further exacerbates the problem. Small sample sizes can lead to overfitting, where the model learns to recognize specific instances rather than general patterns. This overfitting can degrade the model's generalization capabilities, resulting in poorer performance on unseen data. Therefore, in this experiment, the recognition and classification of the “Other” category were excluded. However, future research will place greater emphasis on exploring more sophisticated methods to handle the heterogeneity and limited sample size within the “Other” category. This will involve investigating advanced techniques such as data augmentation, transfer learning, and the development of more robust models capable of generalizing from diverse and rare pathological cases. 
\par In the future, we will continue to explore more advanced techniques to address the challenges posed by the “Other” category in the ViTaL dataset. We believe that through the application of data augmentation, transfer learning, and the development of more robust models, we can improve the model's ability to handle the heterogeneity and limited sample size within this category.
\section{Conclusion}
In this paper, we first collect and annotate a dataset named ViTaL for ovarian tumor recognition, which is the first multimodal multipathology dataset that incorporates text, tables, and other modalities. Subsequently, we introduce ViTaLNet based on the Triplet Hierarchical Offset Attention Mechanism (THOAM) to minimize the loss incurred during feature fusion of multi-modal data. This mechanism could effectively enhance the relevance and complementarity between information from different modalities. Compared with other advanced 2D image recognition classification methods, ViTaLNet achieves relatively better results on the dataset. Moreover, we demonstrate the effectiveness of different components through ablation studies and visualization. Visualization and analysis also prove that ViTaL-Net is convincing and interpretable. I firmly believe that our work holds the potential to bring about substantial value in the clinical recognition of ovarian tumors. By introducing the ViTaL dataset and the ViTaLNet framework, we aim to advance the field of ovarian tumor diagnosis through the integration of multimodal data and innovative classification methods. Our results demonstrate the effectiveness of our approach, and we hope that this work will contribute to more accurate and efficient clinical decision-making processes in the future. 
\par However, our work still has some limitations. In terms of establishing a public dataset: there are fewer samples in some pathological categories, such as High-grade serous carcinoma, etc., which can lead to overfitting and affect the generalization ability of the model. Future work should focus on expanding the dataset to include more samples from underrepresented categories, possibly through multi-center collaborations or the use of data augmentation techniques to artificially increase the diversity of the training data.

\section*{CRediT Author Statement}
\textbf{You Zhou}: Conceptualization, Methodology, Software, Writing - Original Draft. \textbf{Lijiang Chen}: Investigation, Writing - Review \& Editing. \textbf{Guangxia Cui}: Data Curation, Visualization. \textbf{Wenpei Bai}: Funding acquisition. \textbf{Yu Guo}: Data Curation. \textbf{Shuchang Lyu}: Methodology, Writing - Review \& Editing. \textbf{Guangliang Cheng}: Methodology, Writing - Review \& Editing. \textbf{Qi Zhao}: Conceptualization, Project administration.

\section*{Acknowledgment}
All the data collected have approvals from the ethics committees of Beijing Shijitan Hospital, Capital Medical University. This work was supported by the Capital's Funds for Health Improvement and Research (Grant Number: CFH 2024-1-2084).

\bibliographystyle{elsarticle-num} 
\bibliography{ref}

\end{document}